\documentclass[aps,prd,twocolumn,superscriptaddress,nofootinbib,10pt]{revtex4-1}   

\pdfoutput=1
\usepackage[utf8]{inputenc}
\usepackage{amssymb}
\usepackage{amsmath}
\usepackage{amsfonts}
\usepackage{graphicx}
\usepackage{color}
\usepackage{xspace}
\usepackage{comment}
\usepackage{hyperref}
\usepackage[normalem]{ulem}

\usepackage[section]{placeins}
\usepackage{afterpage}

\usepackage{float}
\usepackage{slashed}
\usepackage{ulem}
\usepackage{appendix}

\usepackage{multirow,rotating}
\usepackage[dvipsnames]{xcolor}
\usepackage{comment}

\usepackage{orcidlink}

\begin{document}
	
    \title{Searching for dark photons from dark-scalar decays at CEPC and FCC-ee}

    \affiliation{School of Physics, Hefei University of Technology, Hefei 230601, China}

    \author{Kingman Cheung\,\orcidlink{0000-0003-2176-4053}}
    \email{cheung@phys.nthu.edu.tw}
    \affiliation{Department of Physics, National Tsing Hua University,	Hsinchu 300, Taiwan}
    \affiliation{Center for Theory and Computation, National Tsing Hua University, Hsinchu 300, Taiwan}
    \affiliation{Department of Physics, Konkuk University, Seoul 05029, Republic of Korea}
   
    \author{Fei-Tung Chung\,\orcidlink{0009-0006-5247-3215}}
    \email{feitung.chung@gapp.nthu.edu.tw}
    \affiliation{Department of Physics, National Tsing Hua University,	Hsinchu 300, Taiwan}
   
    \author{Zeren Simon Wang\,\orcidlink{0000-0002-1483-6314}}
    \email{wzs@hfut.edu.cn}
    \affiliation{School of Physics, Hefei University of Technology, Hefei 230601, China}
 
    \begin{abstract}
        We investigate the sensitivity of proposed CEPC and FCC-ee with a center-of-mass energy of 240~GeV to long-lived dark photons heavier than 2~GeV that are pair-produced via the prompt decays of a light scalar mixed with the Standard-Model Higgs boson. We compute the production and decay rates of both the light scalar and the dark photon, and develop two search strategies targeting displaced vertices within the inner tracker of the main detectors. Using Monte Carlo simulations, we evaluate the signal acceptance and projected sensitivity for each strategy. Our results show that, for the scalar-Higgs mixing angle set at $10^{-2}$ just below the current upper limit, the proposed searches at CEPC and FCC-ee can probe dark-photon kinetic-mixing parameter several orders of magnitude below existing bounds, for dark photons lighter than half the dark-scalar mass.
    \end{abstract}

		
	\vskip10mm
	
	\maketitle
	\flushbottom
 
\section{Introduction}\label{sec:intro}

One of the most profound mysteries in modern cosmology and particle physics is the nature of dark matter (DM).
The existence of DM is strongly indicated by multiple astronomical observations, including gravitational lensing, the anisotropies of the cosmic microwave background, and the distribution of galaxies in large-scale surveys.
Despite its overwhelming abundance, the fundamental nature of DM remains elusive.
Various DM candidates have been proposed, ranging from weakly interacting massive particles and axion-like particles to dark gauge bosons, dark fermions, and other exotic particles predicted in extensions of the Standard Model (SM)~\cite{Beacham:2019nyx,Agrawal:2021dbo,Antel:2023hkf}.

In recent years, a class of models called ``portal-physics'' have been put under scrutiny by both theorists and experimentalists, considering both their simple constructions and strong physics motivations.
They are supposed to be the portals that connect the visible sector (the SM spectrum) to a dark sector consisting of the DM, and often include dark photons~\cite{Okun:1982xi,Galison:1983pa,Holdom:1985ag,Boehm:2003hm,Pospelov:2008zw,Curtin:2014cca} or dark scalars~\cite{OConnell:2006rsp,Wells:2008xg,Bird:2004ts,Pospelov:2007mp,Krnjaic:2015mbs,Boiarska:2019jym}.
The dark photon $\gamma'$ is a mediator of a hypothetical, broken dark $U(1)_D$ gauge symmetry that kinetically mixes with the SM photon, via the mixing term $\epsilon F_{\mu\nu} F'^{\mu\nu}$, where $F_{\mu\nu}$ and $F'_{\mu\nu}$ are the field strength tensors of the SM photon and the dark photon, respectively, and $\epsilon$ is the kinetic mixing parameter.
Through the kinetic mixing, the dark photon interacts with the SM particles with a coupling strength proportional to $\epsilon e$, where $e$ is the electromagnetic coupling constant.

The dark scalar or dark Higgs boson, denoted as $\phi$ in this work, is a scalar-portal model predicted in, for instance, scalar-singlet-extensions of the SM.
It interacts with the SM particles via its mixing $\theta$ with the SM Higgs boson $h$.

Furthermore, in some model constructions beyond the SM (BSM), such as the Hidden Abelian Higgs Model~\cite{Wells:2008xg,Gopalakrishna:2008dv}, both the dark scalar and the dark photon are present where the dark photon acquires mass via the Higgs mechanism of the dark scalar.
In this work, we focus on the low-energy phenomenology of this model and investigate, with the technique of Monte Carlo (MC) simulation, the search prospect of long-lived dark photons produced from dark-scalar decays at the next-generation electron-positron colliders, CEPC~\cite{CEPCStudyGroup:2018ghi,CEPCStudyGroup:2023quu,CEPCStudyGroup:2025kmw,Ai:2025cpj} and FCC-ee~\cite{FCC:2018evy}.
We restrict ourselves to the operation mode as the Higgs factories running at the center-of-mass (COM) energy $\sqrt{s}=240$~GeV, where the dark scalar can be singly produced via its mixing with $h$: $e^+ e^- \to Z^*\to Z \,\phi$, and the dark photons are pair produced subsequently by the prompt decay of $\phi$: $\phi\to \gamma'\gamma'$.
To our knowledge, searches for long-lived dark photons in this signal process have not been studied\footnote{See Ref.~\cite{Fuks:2025jrn}, however, for a similar study for the proton-proton colliders.}.
For other phenomenological studies on theoretical scenarios involving both a dark Higgs boson and a dark photon, we refer to, e.g., Refs.~\cite{Wells:2008xg,Ahlers:2008qc,Gopalakrishna:2008dv,Batell:2009yf,Weihs:2011wp,Davoudiasl:2013aya,Curtin:2013fra,Falkowski:2014ffa,Curtin:2014cca,Bakhet:2015pqa,Izaguirre:2018atq,Jodlowski:2019ycu,Araki:2020wkq,Foguel:2022unm,Cheung:2024oxh,Araki:2024uad}.

We confine ourselves to the mass ranges $m_\phi\in [25~\text{GeV}, 140~\text{GeV}]$ and $m_{\gamma'}\in[2~\text{GeV}, m_\phi/2]$.
For these mass ranges, the current bounds on the mixing parameters $\epsilon$ and $\theta$ stem primarily from collider experiments.
For the dark photon, the leading bounds on $\epsilon$ were obtained at the LHC collaborations CMS~\cite{CMS:2019buh} and LHCb~\cite{LHCb:2019vmc}, as well as the $B$-factory experiment BaBar~\cite{BaBar:2014zli}, roughly all in the order of $10^{-3}$.

For $m_\phi$ between 100~GeV and 1~TeV, constraints around $\theta \lesssim \mathcal{O}(10^{-1})$ were achieved by searches at CMS~\cite{CMS:2015hra,CMS:2018amk,CMS:2021yci} and ATLAS~\cite{ATLAS:2018sbw,ATLAS:2020fry,ATLAS:2020tlo,ATLAS:2022xzm}.
For $m_\phi $ between 10~GeV and 100~GeV, searches at CMS~\cite{CMS:2018zvv}, ATLAS~\cite{ATLAS:2021hbr}, LEP2~\cite{LEPWorkingGroupforHiggsbosonsearches:2003ing}, and L3~\cite{Acciarri:1996aaa} have placed upper bounds $\theta \lesssim \mathcal{O}(10^{-2})$.
See Ref.~\cite{Ferber:2023iso} for a detailed summary of these existing bounds.
In this work, we will simply assume $\theta=10^{-2}$ in numerical studies.

For existing studies on the search prospect of dark photons and dark scalars at CEPC and FCC-ee, we refer to, for instance, Refs.~\cite{He:2017zzr,He:2017ord,Biswas:2017lyg,Park:2023ygi,Liu:2017lpo,Inan:2021dir,Jiang:2019hfn,Chang:2018pjp} and Refs.~\cite{Liu:2017lpo,Chang:2018pjp,Cheung:2019qdr,Alipour-Fard:2018lsf,Wang:2019xvx,Ripellino:2024iem,Bhattacherjee:2025dlu,Olgoso:2025jot}, respectively.

We note that there is no published experimental search for the scenario involving both the dark scalar and the dark photon in the mass ranges of our interest.

This paper is organized as follows.
In Sec.~\ref{sec:model}, we introduce our theoretical model.
Then in Sec.~\ref{sec:experiment_search} we discuss the main detectors of CEPC and FCC-ee, and elaborate on our search strategies, including the computation of detector acceptances to the long-lived dark photons.
We proceed in Sec.~\ref{sec:simulation_calculation} to describe the procedures of our MC simulation and final computation of the signal events.
We then present the numerical results in Sec.~\ref{sec:results} and conclude the paper in Sec.~\ref{sec:conclusions}.

\section{Theoretical model}\label{sec:model}

We consider an extension of the SM that introduces an additional dark-sector gauge symmetry $U(1)_D$, under which all SM fields are uncharged.
The new model includes both a dark photon $\gamma'$ and a dark scalar $\phi$.
After spontaneous symmetry breaking of the $U(1)_D$ via the vacuum expectation value (VEV) of the associated dark-scalar field, the dark photon acquires a mass $m_{\gamma'}$ and an interaction between $\phi$ and two dark photons arises.
The strength of this interaction scales with $m_{\gamma'}$ and the $U(1)_D$ gauge coupling $g'$.
After the electroweak symmetry breaking, the dark photon will mix with the SM photon via a gauge kinetic-mixing term linking the SM hypercharge and the new $U(1)_D$ field strength tensors, $\epsilon F_{\mu\nu}F^{'\mu\nu}$.
In addition, the dark scalar $\phi$ mixes with the SM Higgs boson and thus couples to pairs of SM fermions, though the coupling strength is suppressed by the mixing parameter $\theta$.
Thus, the low-energy phenomenological interaction Lagrangian employed in this study takes the form~\cite{Araki:2020wkq}:
\begin{equation}
    L_{\text{int}} = g' m_{\gamma'} \phi \gamma'_\mu \gamma'^\mu + \epsilon \,e \, \gamma'_\mu \, J_{\text{EM}}^\mu + \sum_f \frac{m_f \theta}{v} \phi \bar{f} f, \label{eq:Lagrangian}
\end{equation}
where $e=\sqrt{4\pi\alpha} \approx 0.31$ and $J_{\text{EM}}^\mu$ are the electromagnetic coupling 
constant and the SM electromagnetic current, respectively, with $\alpha$ being the fine-structure constant. 
In the last term, $v=246$~GeV is the VEV of the SM Higgs field 
and the summation goes over all SM fermions.

At the future Higgs factories, the SM Higgs bosons are dominantly produced via the Higgs-strahlung process, $e^+ e^- \to Z\,h$.
The dark scalar is produced in a very similar process, $e^+ e^-\to Z\,\phi$, which is allowed as long as $\phi$ is light enough and it mixes with $h$.
The cross section of the $e^+ e^-\to Z\,\phi$ process is thus suppressed by $\theta^2$, and reads~\cite{Djouadi:2005gi}
\begin{equation}
    \sigma(e^+ e^- \to Z\,\phi) = \frac{G_F^2 m_Z^4}{96\pi s} (v_e^2 + a_e^2) \sqrt{\lambda} \frac{\lambda + 12 m_Z^2 / s}{(1 - m_Z^2 / s)^2} \cdot \theta^2,\label{eqn:phi_prod_xs}
\end{equation}
where $G_F$ is the Fermi constant, $m_Z$ is the $Z$-boson mass, $s$ is the COM energy squared, $v_e = -1 + 4 \sin^2 \theta_w$ and $a_e = -1$ are, respectively, the vector and axial vector $Z$-boson charges of the electron, and $\theta_w$ is the weak mixing angle.
$\lambda$ is defined as
\begin{equation}
    \lambda = \left( 1 - \frac{(m_\phi + m_Z)^2}{s} \right)\left( 1 - \frac{(m_\phi - m_Z)^2}{s} \right).
\end{equation}

\begin{figure}[t]
    \centering
    \includegraphics[width=\linewidth]{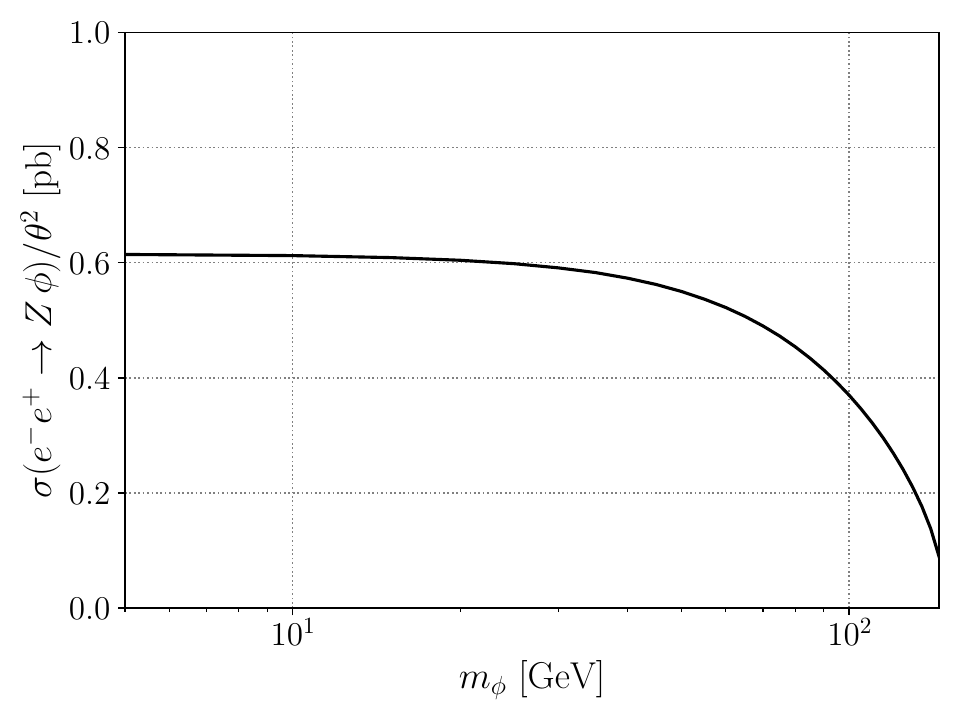}
    \caption{The production cross section of $\phi$ via the process $e^+ e^- \to Z \,\phi$ rescaled with $1/\theta^2$ as a function of the dark-scalar mass $m_\phi$, computed with \texttt{MadGraph5}~\cite{Alwall:2014hca}, at the COM energy $\sqrt{s} = 240$ GeV.}
    \label{fig:cross_ee2Zphi}
\end{figure}

In Fig.~\ref{fig:cross_ee2Zphi} we plot the scattering cross section $\sigma(e^+ e^-\to Z\,\phi)$ rescaled with $1/\theta^2$ at the COM energy $\sqrt{s}=240$~GeV, as a function of $m_\phi$, computed with \texttt{MadGraph5~3.4.2}~\cite{Alwall:2014hca}.
We find these results in agreement with Eq.~\eqref{eqn:phi_prod_xs}.

As mentioned earlier, in this work, we focus on $m_\phi\sim 10\text{--}100$~GeV.
With Eq.~\eqref{eq:Lagrangian}, the partial decay widths of the dark scalar into a pair of dark photons and a pair of lighter SM fermions can be calculated with~\cite{Araki:2020wkq}
\begin{align}
    \Gamma_{\phi \to \gamma' \gamma'} &= \frac{g'^2}{8\pi} \frac{m_{\gamma'}^2}{m_\phi} \left( 2 + \frac{m_\phi^4}{4m_{\gamma'}^4} \left( 1 - \frac{2m_{\gamma'}^2}{m_\phi^2} \right)^2 \right) \beta_\phi(\gamma'), \label{eq:Gamma_phi2gamma'gamma'} \\
    \Gamma_{\phi \to f \bar{f}} &= N_c \frac{m_\phi}{8\pi} \left( \frac{m_f}{v} \right)^2 \theta^2 \left( 1 - \frac{4m_f^2}{m_\phi^2} \right) \beta_\phi(f), \label{eq:Gamma_phi2ff}
\end{align}
where $\beta_i(j) \equiv \sqrt{1 - 4m_j^2/m_i^2}$ is the phase-space factor for the two-body decay $i \to jj$, $N_c = 1 \, (3)$ for $f$ being a lepton (quark).

The partial decay widths of $\phi \to f \bar{f}$ are suppressed by $\theta^2$ (see Eq.~\eqref{eq:Gamma_phi2ff}) while that of $\phi\to \gamma'\gamma'$ is controlled by $g'$ which is, at present, only loosely constrained by theoretical arguments such as perturbative unitarity.
In this work, we fix $g'$ at $10^{-2}$ so that not only the theoretical constraints are satisfied but also the dark scalar $\phi$ dominantly decays into a pair of dark photons as calculated with Eq.~\eqref{eq:Gamma_phi2gamma'gamma'} and Eq.~\eqref{eq:Gamma_phi2ff} (note that we fix $\theta$ at $10^{-2}$).
Further, calculation with Eq.~\eqref{eq:Gamma_phi2gamma'gamma'} shows that $\phi$ decays promptly.
We thus safely assume that $\phi$, once produced at the interaction point (IP), decays at the same position, into a pair of (long-lived) dark photons.

The dark photon decays into a pair of SM charged fermions, with the partial decay widths~\cite{Bauer:2018onh,Batell:2009yf,Araki:2020wkq,DOnofrio:2019dcp,Fabbrichesi:2020wbt,Cheung:2024oxh}
\begin{equation}
    \Gamma_{\gamma' \to f \bar{f}} = \frac{1}{3} N_c \, Q_f^2 \, \alpha \, \epsilon^2 \, m_{\gamma'} \, \left( 1 + \frac{2m_f^2}{m_{\gamma'}^2} \right) \, \beta_{\gamma'}(f), \label{eq:Gamma_gamma'2ff}
\end{equation}
where $Q_f$ and $m_f$ are, respectively, the electric charge and the mass of the produced fermion.
We note that this formula is valid in the perturbative regime, $m_{\gamma'} \gtrsim 2$~GeV, as is the case in this study.

\begin{figure}[t]
    \centering
    \includegraphics[width=0.99\linewidth]{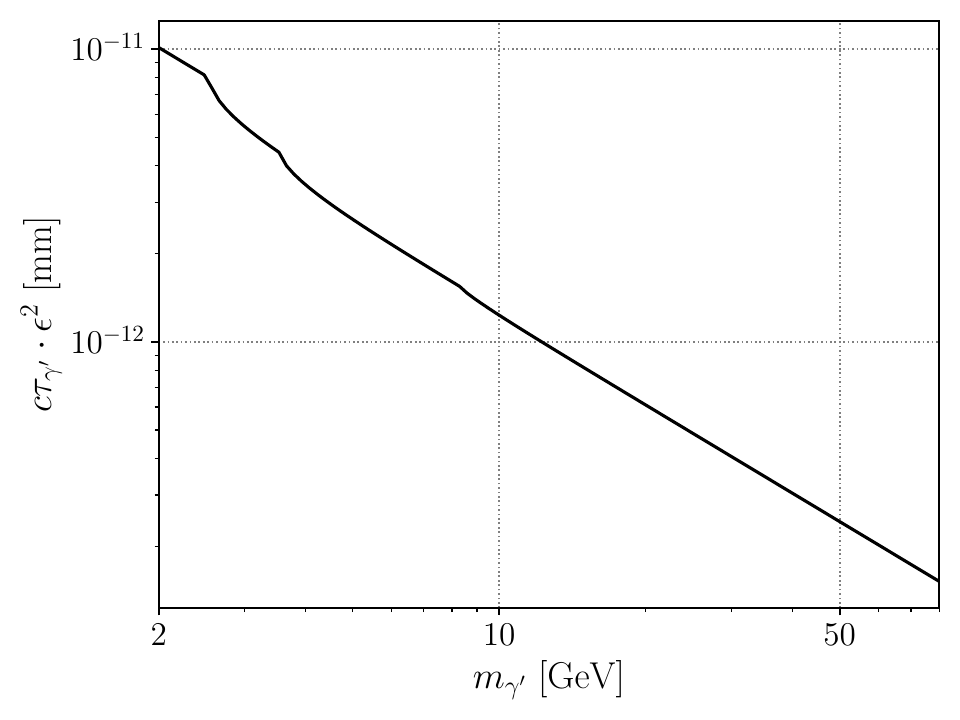}
    \caption{The proper decay length $c\tau_{\gamma'}$ of the dark photon, rescaled with $\epsilon^2$, as a function of the dark-photon mass.}
    \label{fig:ctau_rescaled}
\end{figure}

In Fig.~\ref{fig:ctau_rescaled}, we plot $c \tau_{\gamma'} \cdot \epsilon^2$, the $\epsilon^2$-rescaled proper decay length of the dark photon, as a function of $m_{\gamma'}$.
Here, $c\tau_{\gamma'}=c\hbar/\Gamma_\text{total}$, with $c$ and $\hbar$ denoting the speed of light and the reduced Planck constant, respectively, and $\Gamma_\text{total}$ the total decay width of $\gamma'$.
For sufficiently small $\epsilon$ or $m_\gamma'$, and a sufficiently large Lorentz boost, the dark photon becomes long-lived.

\begin{figure}[t]
    \centering
    \includegraphics[width=0.99\linewidth]{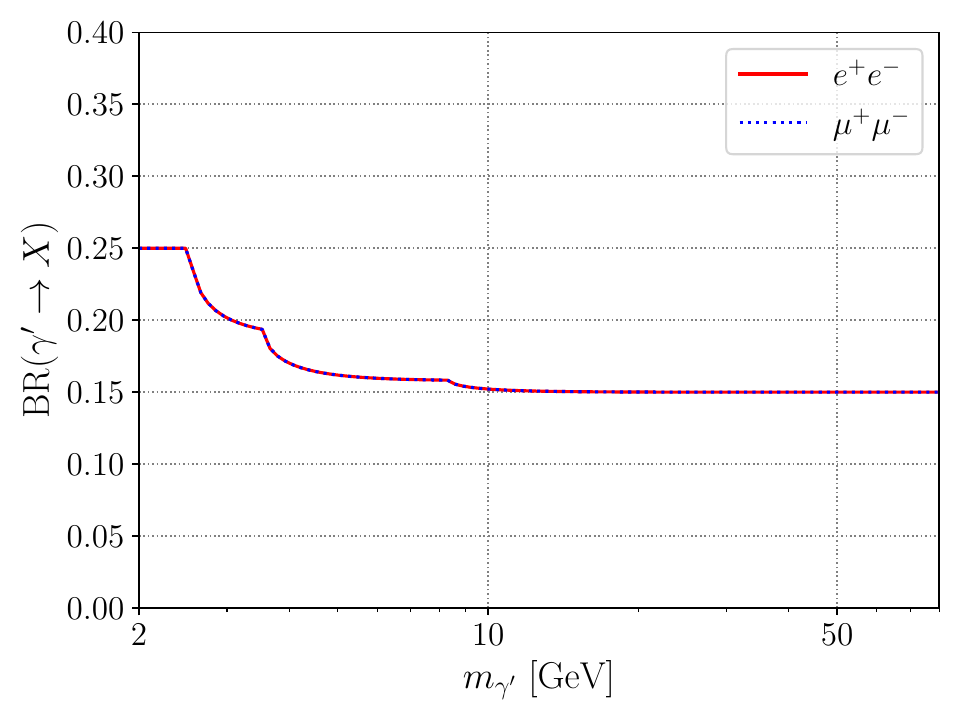}
    \caption{The decay branching ratios of the dark photon into $e^+e^-$ (red solid) and $\mu^+\mu-$ (blue dotted), respectively, as functions of the dark-photon mass.
    Note that the two curves overlap for the whole range of $m_{\gamma'}$ shown here.
    The kinks at $m_{\gamma'}\sim 2.5$~GeV, 3.5~GeV, and 8.3~GeV, arise from the opening of the dark-photon decay channels into a pair of charm quarks, $\tau$-leptons, and bottom quarks, respectively.}
    \label{fig:br2ee_br2mumu}
\end{figure}
Further, since in the search strategies to be discussed in Sec.~\ref{sec:experiment_search} we will consider the final states of an electron-pair or a muon-pair from each dark-photon decay, we show in Fig.~\ref{fig:br2ee_br2mumu} the decay branching ratios (BRs) of the dark photon into these particles as functions of $m_{\gamma'}$.
Since the electron and the muon share the same charge and both of their masses are much lighter than the dark-photon mass values we cover, the two BRs overlap with each other. 
The remaining decay modes are into tau leptons and hadronic final states.

\begin{figure}[t]
    \centering
    \includegraphics[width=0.96\linewidth]{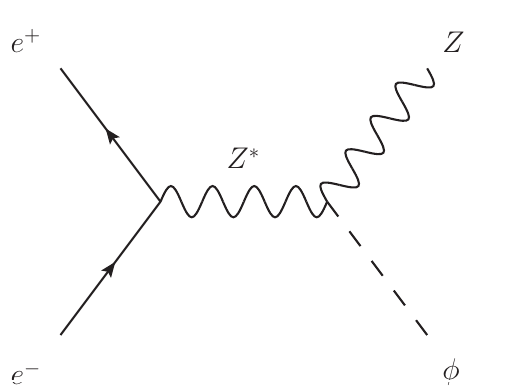}
    \includegraphics[width=0.96\linewidth]{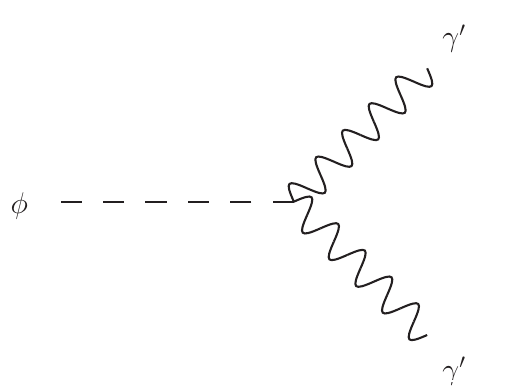}
    \caption{The Feynman diagrams for production (upper plot) and decay (lower plot) of the dark scalar.}
    \label{fig:feynman}
\end{figure}

In addition, we provide in Fig.~\ref{fig:feynman} the Feynman diagrams illustrating the production and decay processes of the dark scalar.

\section{The experiments and search strategies}\label{sec:experiment_search}

\begin{figure}[t]
    \centering
    \includegraphics[width=0.99\linewidth]{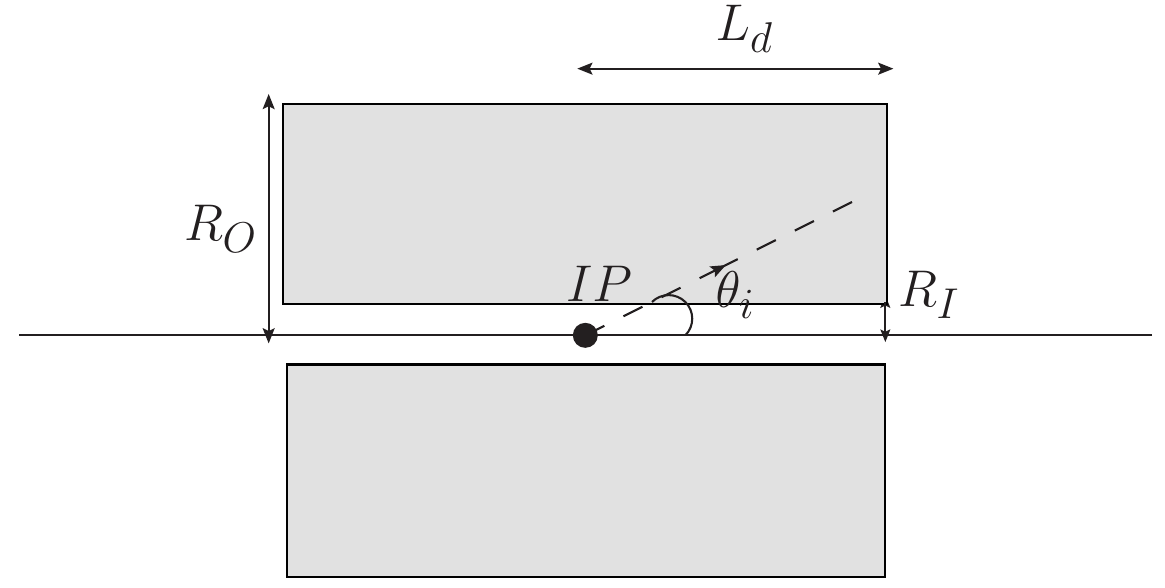}
    \caption{A profile sketch of the inner tracker at CEPC and FCC-ee, extracted from Ref.~\cite{Cheung:2019qdr}.
    }
    \label{fig:detector_structure}
\end{figure}

The main detectors of CEPC and FCC-ee share similar components and geometrical structures, while their exact dimensions differ to some extent.
In particular, we will focus on the inner tracker (IT) component of the main detectors in our search strategies.
We thus show a profile sketch of the IT at the CEPC or FCC-ee main detector in Fig.~\ref{fig:detector_structure}, reproduced from Ref.~\cite{Cheung:2019qdr}.
In Fig.~\ref{fig:detector_structure}, we denote a long-lived particle $i$ with polar angle $\theta_i$ with a dashed line.
$R_I$ ($R_O$) is the inner (outer) radius of the IT, and $L_d$ is the half length of the IT.

\begin{table}[t]
    \centering
    \begin{tabular}{c|c|c|c|c}
        \hline
        Detector & $R_I$ [mm] & $R_O$ [m] & $L_d$ [m] & $V$ [m$^3$] \\
        \hline
        \hline
        CEPC & 16 & 1.8 & 2.35 & 47.8 \\
        \hline
        FCC-ee & 17 & 2.0 & 2.0 & 50.3 \\
        \hline
    \end{tabular}
    \caption{Parameters of the inner tracker at CEPC and FCC-ee, extracted from Ref.~\cite{CEPCStudyGroup:2018ghi} and Ref.~\cite{FCC:2018evy}, respectively.}
    \label{tab:parameters_inner}
\end{table}

Further, we list in Table~\ref{tab:parameters_inner} the geometrical parameters of the inner trackers at CEPC~\cite{CEPCStudyGroup:2018ghi} and FCC-ee~\cite{FCC:2018evy}.
Here, $V$ denotes the volume.

We focus on the case where the dark photons are long-lived, so that their decay positions are displaced from the IP.
We employ two search strategies based on displaced vertices (DVs) within the IT:
\begin{enumerate}
    \item At least one DV consisting of an electron pair or a muon pair should be reconstructed.
    \item Exactly two DVs should be reconstructed, composed of any final states of the dark-photon decays.
\end{enumerate}
The portion of signal events satisfying the requirements given above, i.e.~the acceptance $\mathcal{A}$, can be estimated with the following formulas,
\begin{align}
    \mathcal{A}_1 &= \frac{1}{N_{\text{MC}}} \sum_{i=1}^{N_{\text{MC}}} \bigg\{ P(\gamma'_1 \text{ in IT}) \cdot \Big( 1 - P(\gamma'_2 \text{ in IT}) \Big) \cdot \mathcal{B} \nonumber \\
    &\phantom{{}={}} + P(\gamma'_2 \text{ in IT}) \cdot \Big( 1 - P(\gamma'_1 \text{ in IT}) \Big) \cdot \mathcal{B} \nonumber \\
    &\phantom{{}={}} + P(\gamma'_1 \text{ in IT}) \cdot P(\gamma'_2 \text{ in IT}) \cdot \mathcal{B}^2 \bigg\}, \label{eq:A1}\\
    \mathcal{A}_2 &= \frac{1}{N_{\text{MC}}} \sum_{i=1}^{N_{\text{MC}}} \bigg\{ P(\gamma'_1 \text{ in IT}) \cdot P(\gamma'_2 \text{ in IT}) \bigg\}, \label{eq:A2}\\
    \mathcal{B} &\equiv \text{BR}(\gamma' \to e^+e^-) + \text{BR}(\gamma' \to \mu^+\mu^-),  \label{eq:BR}
\end{align}
where $\mathcal{A}_{1/2}$ denotes the acceptance of the first/second strategy, $N_{\text{MC}}$ is the total number of the MC simulated events, and $\gamma'_{1/2}$ labels the first/second dark photon in each simulated event.
Further, the function $P(\gamma'\text{ in IT})$ is for computing the probability of $\gamma'$ decaying inside the IT, taking the inputs of the kinematics and lifetime of $\gamma'$ as well as the geometries of 
the IT, and is given below,
\begin{align}
    P(\gamma' \text{ in IT}) &= 
    \begin{cases}
        e^{-L_1/\lambda_\text{lab}} - e^{-L_2/\lambda_\text{lab}}, \text{ if } |L_d \tan \theta| > R_I \\
        0, \text{ else,}
    \end{cases} \label{eq:probability_inner} \\
    L_1 &\equiv R_I, \nonumber \\
    L_2 &\equiv \min(|L_d \tan \theta|, R_O), \nonumber \\
    \lambda_\text{lab} &\equiv \beta_T \, \gamma \, c\, \tau_{\gamma'}. \nonumber
\end{align}
Here, $\beta_T$ is the transverse speed of the dark photon with its corresponding Lorentz boost factor labeled as $\gamma$, and $\tau_{\gamma'}$ is its proper lifetime.

Additionally, we have tested one search strategy utilizing the muon chamber:
\begin{itemize}
    \item Two DVs both composed of a muon pair should be reconstructed in the muon chamber.
\end{itemize}
However, our simulation and calculation showed that this strategy has negligible sensitivities to the long-lived dark photons, mainly owing to too low acceptance rates in all the relevant parameter points.
Therefore, we will not discuss this strategy further.

Before discussing our simulation procedures, we briefly comment on potential backgrounds.
For events with two DVs inside the fiducial volume, we expect that the requirement of two spatially separated DVs eliminates all SM backgrounds to a negligible level, independently of the final state of the dark photons.
For events with only one DV, the dominant background for displaced leptons arise from photon conversions, i.e.~photons originating at the IP converting to an $e^+ e^-$ pair on detector materials (most prominently at the beam pipe, which roughly coincides with the inner radius of the IT).
This background is strongly suppressed by our choice of fiducial volume (see Eq.~\eqref{eq:probability_inner}), and can be further reduced by using a detailed material map and by requiring that the DV direction should not be collinear with any reconstructed prompt photon.
Other potential backgrounds include hadronic secondary interactions (tending to produce multi-track vertices), $K_S\to \pi^+\pi^-$ or heavy-flavor decays with pion-to-lepton misidentification, and random track combinatorics.
Cuts on track multiplicity, lepton identification, invariant mass of the dilepton, and DV pointing reduce these sources to a negligible level.
Given these considerations, we assume a background-free environment for our phenomenological sensitivity estimates.

\section{Simulation and calculation}\label{sec:simulation_calculation}

We perform MC simulations with the tool \texttt{Pythia8.3}~\cite{Bierlich:2022pfr}, scanning over $m_{\gamma'}$ and $\tau_{\gamma'}$ for representative values of $m_\phi$, in order to estimate the number of signal events $N_S$.
Concretely, we select the following four values of $m_\phi$: 25, 50, 90, and 140~GeV, and for each value of $m_\phi$, we scan over $m_{\gamma'}$ from 2~GeV to $m_\phi/2$.
Specifically, at every scanned grid point of $(m_{\phi}, m_{\gamma'}, \tau_{\gamma'})$, we simulate ten thousand signal events.
We then translate each combination of $(m_{\gamma'}, \tau_{\gamma'})$ into $(m_{\gamma'},\epsilon)$ with the help of Eq.~\eqref{eq:Gamma_gamma'2ff} and the relation $\Gamma_{\text{total}} = \displaystyle\sum_f \Gamma_{\gamma' \to f \bar{f}} = \hbar / \tau_{\gamma'}$.

In \texttt{Pythia8}, we turn on the process \texttt{HiggsSM:ffbar2HZ} at a COM energy $\sqrt{s}=240$~GeV with back-to-back electron-positron collisions.
By tuning the mass of the SM Higgs boson $h$ to match that of the dark scalar $\phi$, we mimic the kinematics of the signal process $e^+e^-\to Z\,\phi$.
The dark scalar $\phi$ is then set to decay promptly and exclusively into a pair of long-lived dark photons.
The kinematics of the dark photons in each signal event are then used as input for calculating the acceptance $\mathcal{A}_i$ of each search strategy; see Eq.~\eqref{eq:A1}--Eq.~\eqref{eq:probability_inner}.

Finally, we compute the number of signal events with
\begin{equation}
    N_{S,i} = \mathcal{L}_\text{int.} \cdot \sigma(e^+ e^- \to Z\,\phi) \cdot \mathcal{A}_i,
\end{equation}
with $i = 1, 2$ for the two search strategies.
$\sigma(e^+ e^- \to Z\,\phi)$ is computed with \texttt{MadGraph5}; see Fig.~\ref{fig:cross_ee2Zphi} and the relevant discussion in Sec.~\ref{sec:model}.
Here, $\mathcal{L}_\text{int.}$ denotes the integrated luminosity, with $\mathcal{L}_\text{int.} = 22$~ab$^{-1}$ at CEPC~\cite{Ai:2025cpj} and $\mathcal{L}_\text{int.} = 5.6$~ab$^{-1}$ at FCC-ee~\cite{FCC:2018evy}.

For simplicity, we assume a detector efficiency of 100\% in the calculation.
Additionally, the reconstruction efficiency of DVs, in principle, degrades with increasing distances from the IP.
In particular, at positions close to the outer edge of the tracker we expect next to negligible DV reconstruction efficiencies.
In expectation of future dedicated improvement on the experimental techniques, hardware, and search strategies, we choose to neglect such degrading effects and take an optimistic flat efficiency of 100\% for the DV reconstruction.

\section{Sensitivity reach}\label{sec:results}

\begin{figure}[t]
    \centering
    \includegraphics[width=0.99\linewidth]{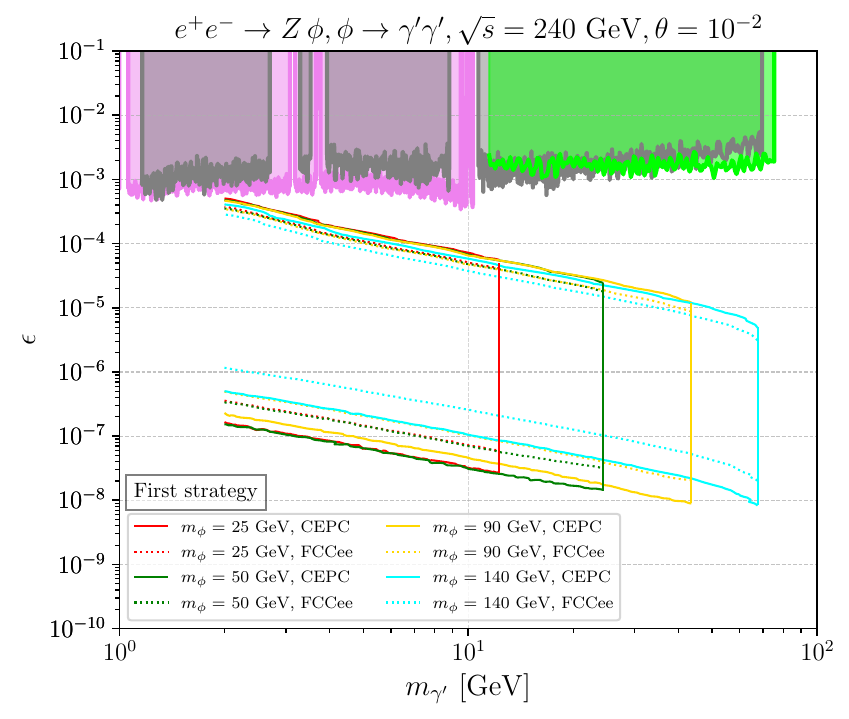}
    \includegraphics[width=0.99\linewidth]{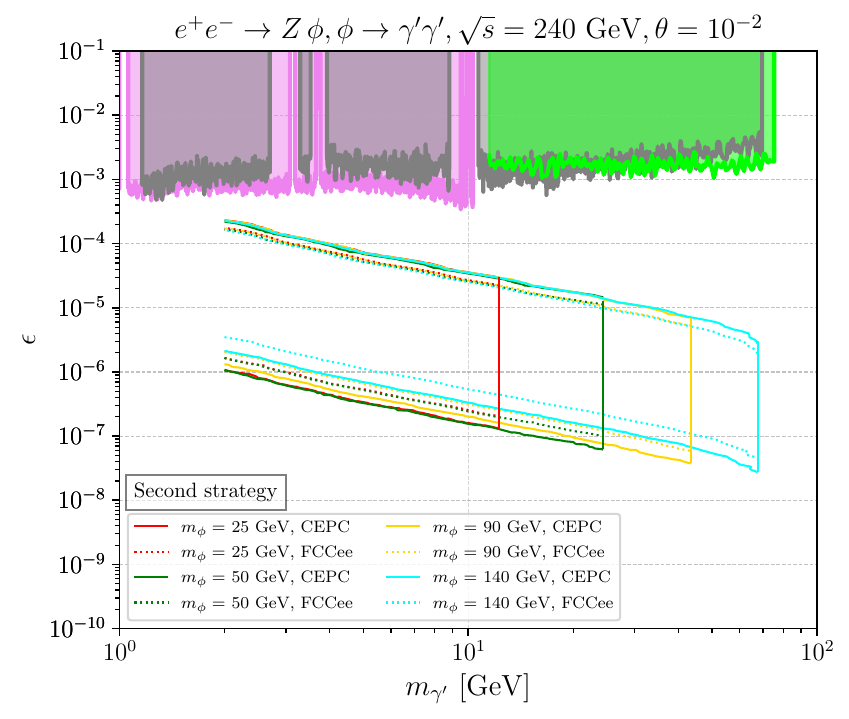}
    \caption{Sensitivity results of the first (upper panel) and second (lower panel) strategies.
    Here, the COM energy is set at $240$~GeV and the scalar-Higgs mixing angle is fixed at $10^{-2}$.
    The gray-, green-, and pink-filled regions have been excluded by searches at LHCb~\cite{LHCb:2019vmc}, CMS~\cite{CMS:2019buh}, and BaBar~\cite{BaBar:2014zli}, respectively.}
    \label{fig:result_strategy}
\end{figure}

In this section, we present the numerical results.
As discussed above, our search analyses are essentially background-free.
We thus show 3-signal-event isocurves as the exclusion bounds at 95\%~confidence level (C.L.) for vanishing background events.
We display the sensitivity results in the ($m_{\gamma'}$, $\epsilon$) plane, in the two plots of Fig.~\ref{fig:result_strategy} for the two search analyses, respectively.

In Fig.~\ref{fig:result_strategy}, the isocurves of different colors correspond to different values of $m_\phi$, and the solid and dashed line styles are for CEPC and FCC-ee results, respectively.
The parameter regions enclosed by the isocurves can be excluded at 95\% C.L.~if no event is observed.
Above the upper parts of the isocurves, the dark photons are too short-lived to decay inside the fiducial volume of the IT, while below the lower parts, the dark photons are so long-lived that they decay only after leaving the IT.
Further, the upper reach of $m_{\gamma'}$ is determined by the kinematic threshold $m_{\gamma'}<m_\phi/2$.
We note that the curves start at $m_{\gamma'}=2$~GeV, since we choose to consider only dark photons heavier than that.

The existing bounds from searches at LHCb~\cite{LHCb:2019vmc}, CMS~\cite{CMS:2019buh}, and BaBar~\cite{BaBar:2014zli} are shown in the figures as color-filled areas.

These results indicate that with a COM energy of 240~GeV, both CEPC and FCC-ee can test the kinetic-mixing parameter $\epsilon$ several orders of magnitude below the existing limits, with the scalar-Higgs mixing angle $\theta$ fixed at $10^{-2}$.
In the lower parts of the sensitivity curves, we observe that CEPC can probe $\epsilon$ values roughly a factor of 2 ($\sqrt{2}$) smaller than those that FCC-ee can for the first (second) search strategy, primarily owing to its integrated luminosity being larger by about a factor of 4.

Further, compared to the first search strategy, the second one only results in weaker sensitivities.
This is because the sensitivity of the first strategy is dominantly contributed by events with exactly one DV in the fiducial volume, while that of the second one comes solely from events with two DVs inside (as the strategy requires by definition).
The latter is thus doubly suppressed by the exponential decay distribution functions, in contrast to the single exponential suppression in the former case.

In addition, we comment that for $\theta$ values smaller than $10^{-2}$, worse sensitivity reach is expected, because fewer dark scalars and thus fewer dark photons would be produced.
Quantitatively speaking, if $\theta$ is reduced by a factor of, say, 10, the total dark-photon production rates would decrease by 100.
For the first search strategy, the lower sensitivity reach to $\epsilon$ would be weakened by a factor of 10 while that for the second search would be impaired by about a factor of $\sqrt{10}$.

Finally, we note that $\Gamma(\phi\to\gamma'\gamma')$ decreases for smaller values of $g'$ (and for heavier dark photons); see Eq.~\eqref{eq:Gamma_phi2gamma'gamma'}.
Therefore, while we fix both $\theta$ and $g'$ at $10^{-2}$ for numerical studies in this work thus concluding that BR$(\phi\to \gamma'\gamma')\simeq 100\%$ for the whole parameter region of our interest, for an even more suppressed $g'$ coupling it can happen that $\phi$ does not decay exclusively into a pair of dark photons.
In this case, inferior sensitivities are expected.
Also, if both $\theta$ and $g'$ are multiple orders of magnitude smaller, $\phi$ itself can become long-lived like the dark photons, making the collider phenomenology more complicated and most likely weakening the sensitivity reach; see, for example, Ref.~\cite{Wang:2024mrc} for an illustration of this effect.

\section{Conclusions}\label{sec:conclusions}

In this paper, we have proposed to search for long-lived dark photons from dark-scalar decays at CEPC and FCC-ee.
We focus on the signal process $e^+ e^-\to Z\,\phi$ at $\sqrt{s}=240$~GeV, where the dark scalar $\phi$ decays promptly into a pair of the dark photons with a branching ratio of essentially $100\%$.
The dark photons are long-lived for sufficiently small values of kinetic mixing and their masses, decaying into various SM fermion pairs at the parton level.
We employ two DV-based search strategies with different requirements on the decay-product types and the number of DVs inside the IT of the main detectors at CEPC and FCC-ee.

We have computed the production and decay rates of both the dark scalar and the dark photon, and performed MC simulations with \texttt{Pythia8}, scanning over the dark-scalar mass~$m_\phi$, the dark-photon mass~$m_{\gamma'}$, and the dark-photon proper lifetime~$\tau_{\gamma'}$.
We have thus estimated the signal-event rates of the two search strategies, with integrated luminosities of $\mathcal{L}_\text{int.} = 22$~ab$^{-1}$ for CEPC and $\mathcal{L}_\text{int.} = 5.6$~ab$^{-1}$ for FCC-ee, respectively.

Under the legitimate assumption of vanishing background, we present the sensitivity reach of our proposed search analyses to the long-lived dark photons at CEPC and FCC-ee, shown in the $(m_{\gamma'}, \epsilon)$ plane, where we have fixed $\theta$ at the present upper bound $\sim 10^{-2}$.
The numerical results indicate that these search strategies can probe values of $\epsilon$ several orders of magnitude below the existing bounds, for $m_\phi\sim 10\text{ -- }100$~GeV and $m_{\gamma'}$ between 2~GeV and $m_\phi/2$.

We close the paper by commenting briefly on the impact on the sensitivities if a more realistic, detector-level simulation is conducted than the approach applied in the present paper.
In that case, the detector-level factors including particle-identification efficiencies, DV-reconstruction efficiencies, and smearing effects should be properly taken into account, leading to, we expect, an order of $10\%$ reduction on the signal-event numbers\footnote{We quote the 10\% reduction as an order-of-magnitude estimate, inspired by attempts to recast LHC searches for long-lived particles, which often rely on parameterized efficiency functions or efficiency maps to include realistic detector effects. Such functions commonly return efficiency values of this order. See, e.g., Ref.~\cite{Cheung:2024qve}, where some of us recast a recent ATLAS search for displaced vertices and multiple jets~\cite{ATLAS:2023oti}.}, in most of the parameter regions and phase space.
In particular, if the dark photon is only slightly below the kinematic threshold, i.e.~$m_{\gamma'}\lesssim m_{\phi}/2$, the produced dark photons should be much less energetic and its decay products should be rather soft; in this case we expect weakened detectability in general.
However, dedicated triggers and machine-learning techniques, among other possibilities, might help enhance the sensitivities; see, e.g.,~Ref.~\cite{Blekman:2020hwr} for a relevant phenomenological study on soft displaced leptons at the LHC.\footnote{See also Ref.~\cite{Fukuda:2019kbp} for a novel search strategy for a soft microdisplaced track at the LHC.}
Similar search strategies for CEPC and FCC-ee may be inspired from such studies.

\section*{Acknowledgements}
We thank C.J.~Ouseph for useful discussions and Huayang Wang for the participation in the initial stage of the project.
K.C.~is supported by the National Science \& Technology Council (NSTC) of Taiwan under grant no.~NSTC 113-2112-M-007-041-MY3. 
Z.S.W.~is supported by the National Natural Science Foundation of China under grant Nos.~12475106 and 12505120, and the Fundamental Research Funds for the Central Universities under Grant No.~JZ2025HGTG0252.

\bibliography{bib}

\end{document}